# A COMPARATIVE STUDY OF PROCESS MEDIATOR COMPONENTS THAT SUPPORT BEHAVIORAL INCOMPATIBILITY

Kanmani Munusamy[1], Harihodin Selamat[2], Suhaimi Ibrahim[3] and Mohd Sapiyan Baba[4]

[1, 2, 3] Advanced Informatics School (AIS), Universiti Teknologi Malaysia (UTM), Malaysia.
[1]kanmani3@live.utm.my, [2]harihodin@utm.my, [3]suhaimiibrahim@utm.my
[4]Faculty of Computer Science & Information Technology, University Malaya (UM), Malaysia.
[4]pian@um.edu.my

## ABSTRACT

*Most businesses these days use the web services technology as a medium to allow interaction between a service provider and a service requestor. However, both the service provider and the requestor would be unable to achieve their business goals when there are miscommunications between their processes. This research focuses on the process incompatibility between the web services and the way to automatically resolve them by using a process mediator. This paper presents an overview of the behavioral incompatibility between web services and the overview of process mediation in order to resolve the complications faced due to the incompatibility. Several state-of the-art approaches have been selected and analyzed to understand the existing process mediation components. This paper aims to provide a valuable gap analysis that identifies the important research areas in process mediation that have yet to be fully explored.*

## KEYWORDS

*Web services, semantic web services, web service mediation, web service adaptation and behavioral incompatibility*

## 1. INTRODUCTION

There are many recent research studies on the methods to specify (in formal and concise language), compose (automatically), discover, secure, and ensure the correctness of web services [1]. One of the essential issues in web services is behavioral incompatibility which handles heterogeneity, resolves mismatches and establishes interoperability between the web services. There are many terminologies such as interoperability, mismatches, incompatibility and heterogeneity issues that are closely related to mediation and which have been widely discussed in system integration initiatives.

These mismatches occur in the environment of the web services due to the significant increase in the number of web services and the distributive nature of the web services themselves. There are three levels of mediation that have emerged and these are firstly the data mediation to address *signature* level mismatches; secondly the process mediation to address *protocol* level mismatches; and finally the functional mediation which refers to levels of *concept* mismatches in Component-Based Software Engineering (CBSE). However, this paper only focuses on the process mediation in the web services. Therefore, process mediation which centers on the behavioral and communication mismatches during web service interaction is essential in ensuring the compatibility of the web services. The process mediation can be visualized as a middle service that ensures two web services are able to interact successfully.

The previous work on the evaluation of the process mediation approaches focused on the four important evaluation criteria such as expressiveness, automation, correctness and completeness [2]. In this paper, the important components of the process mediation in web service environment have been identified and a comparative study of the existing approaches based on these components has been discussed.

The remaining part of this paper is organized as follows: Section 2 describes the process mediation in a web service and it provides the definition for behavioral incompatibility and the related mismatches. The definition for adaptation and mediation which is widely used in resolving behavioral mismatches is also provided in this section. Section 3 presents the Systematic Literature Review Technique that has been used in the selection of the state-of-the-art approaches while Section 4, provides an overview of process mediation which is followed by the main discussion of this paper. Section 5 provides a comparative evaluation of the current approaches based on a set of evaluation criteria that is related to the process mediation component which is specified in Section 4. Section 6 then provides a summary of the comparative evaluation and discusses the gap analysis in the area of process mediation in the web services. Finally, Section 7 concludes this paper.

## 2. BEHAVIORAL COMPATIBILITY IN WEB SERVICES

Generally, during the discovery and selection phases, the capability of an offered service will meet with the goal of the requested service from various perspectives in order to discover and select the services that are compatible as explained in [3]. However, there is no assurance that these matched services can interact well since they may meet with a deadlock which can only be detected during the actual invocation phase. Behavioral incompatibility in the selected web services can lead to the termination of a web service composition and invocation if they are not detected before the actual execution. Therefore it is important to identify behavioral compatibility between the web services and support the incompatibility using behavioral resolutions such as a mediator or an adaptor.

Behavioral compatibility is mainly concerned with mismatches in the web services interaction. It occurs when a web service is in use or when there's communication between several web services to produce the desired output based on the input received. It also ensures that business process of each individual web services executed correctly. Therefore, behavioral compatibility analysis closely related to business process and rule management techniques as explained in [4].

The research on this behavioral incompatibility has been stated in various terminology such as mismatch in the Message Exchange Pattern (MEP) [5] or behavioral interface, process incompatibility [6] and protocol mismatches [7]. Behavioral compatibility analysis is able to provide answers to questions listed below.

a) Are the matched and selected web services able to communicate successfully?
b) What types of mismatches contribute to unsuccessful communication?
c) Are the identified mismatches solvable?
d) What is the solution for each identified mismatch?

### 2.1 Mismatches in Behavioral Analysis

This section provides the types of mismatches that are related to behavioral incompatibility and their hierarchies. Generally there are four type of mismatches that contribute to the behavioral incompatibility in a web service interaction namely, signature/data, functional, protocol and deadlock.

A signature level of mismatch refers to incompatibility in the data element of a web service with the data element of the interacting web service. The differences in structure, type and naming conventions of the data elements in web service messages can lead to behavioral incompatibility. Generally, the signature level of mismatches uses scheme mapping and data mapping methods to align the differences in the data elements. It requires the involvement of the developer to identify the correct data mappings between the different web service applications.

Next, the functional mismatches describe the differences in the offered and the requested functionalities between the service provider and the requestor. According to [8] there are five types of possible relationships between the offered and requested functionalities, namely equal, plug-in, subsume, intersecting and disjoint. Both of the signature/data and functional level mismatches need to be combined in order to identify the subsequent, protocol level mismatches. Moreover, these levels of mismatches need to be supported by sufficient semantic in order to identify the actual meaning of the interacting messages depending on the web service domain.

The role of semantics in resolving signature/data and functional level mismatches is very essential to ensure the correctness of interaction between the web services. Ontology plays an important role in bringing the semantics in the behavioral analysis. The reasoning mechanism that supports ontology has become very useful in many automated tasks in web services such as discovery, selection and composition. Many Semantic Web Service Frameworks such as OWL-based web service (OWL-S) [9] , Web Service Modeling Ontology (WSMO) [10] and Semantic Annotations of WSDL (SAWSDL) have taken bottom-up approaches to bring the semantics into web services. These approaches specify the goal and capability of the web services semantically from the beginning. This bottom-up method is able to enhance the protocol and deadlock mismatch analysis through the available ontology domain.

The third level of mismatch is called the protocol level. Many researchers like [6, 11, 12] have classified protocol level mismatches into five different types as listed below and these mismatches have been named for reader convenience.

Table 1. List of Protocol Mismatches

| **No** | **Protocol Mismatches** | **Descriptions** |
|---|---|---|
| 1 | Extra Messages (EM) | The sender web service sends messages that are not expected by the receiver. |
| 2 | Missing Messages (MM) | The sender is not sending the messages that are expected by the receiver. |
| 3 | One to Many Messages (OMM) | The sender sends single messages that are expected to be in multiple forms. |
| 4 | Many to One Message (MOM) | The sender sends multiple messages that are expected to be in single form. |
| 5 | Wrong Order Messages (WOM) | The sender sends messages in the wrong order. |

Many other researchers like [11, 13-18] have analyzed these protocol level mismatches with different names such as process mismatch, choreography mismatches and service level mismatches. Finally, the deadlock mismatches refer to the interaction between the web services that are unable to reach the final state or end point. According to [11], the web services that are unable to find the possible mapping due to missing and extra messages can possibly lead to a deadlock situation and [19] defines deadlock as non-final state and for which no outgoing transition exists.

## 2.2 Behavioral Analysis Definition in Web Service

There are two main descriptions of behavioral compatibility that have been adopted in this paper, by [6] and [20]. According to [6], behavioral incompatibility can be classified into three categories namely, precise match, solvable and unsolvable message mismatches. [20] has defined that the two web services are behaviorally compatible only when they have opposite behaviors, no unspecified receptions and are fully free from any deadlock. Generally, both of these definitions have highlighted the common ways to detect behavioral mismatches in web services.

In the first classification of [6], the precise match happens when the sender web service sends the messages in the exact order that the receiver web service requested. This exact mismatch can be identified when both the web services have the opposite behavior as mentioned by [20]. However, the web service interactions that have precise behavioral matches do not assure successful communication due to the incompatibility in their signature and functional level. Therefore data and functional mismatches analyses are very essential in order to ensure the successful interaction in web services.

The second classification on solvable mismatches is similar to the five protocol mismatches that are explained in Section 2.1. These mismatches generally act as the unspecified receptions in the web services interaction. Although these mismatches have been named as solvable mismatches, little evidence shows these mismatches can be detected and resolved successfully without a certain degree of intervention by the developers. Therefore, these mismatches will be termed "protocol mismatches" hence in this paper. Finally, the unsolvable mismatch refers to the deadlock level mismatch. This deadlock level mismatch can be contributed by the other levels of mismatches such as data, functional and protocol.

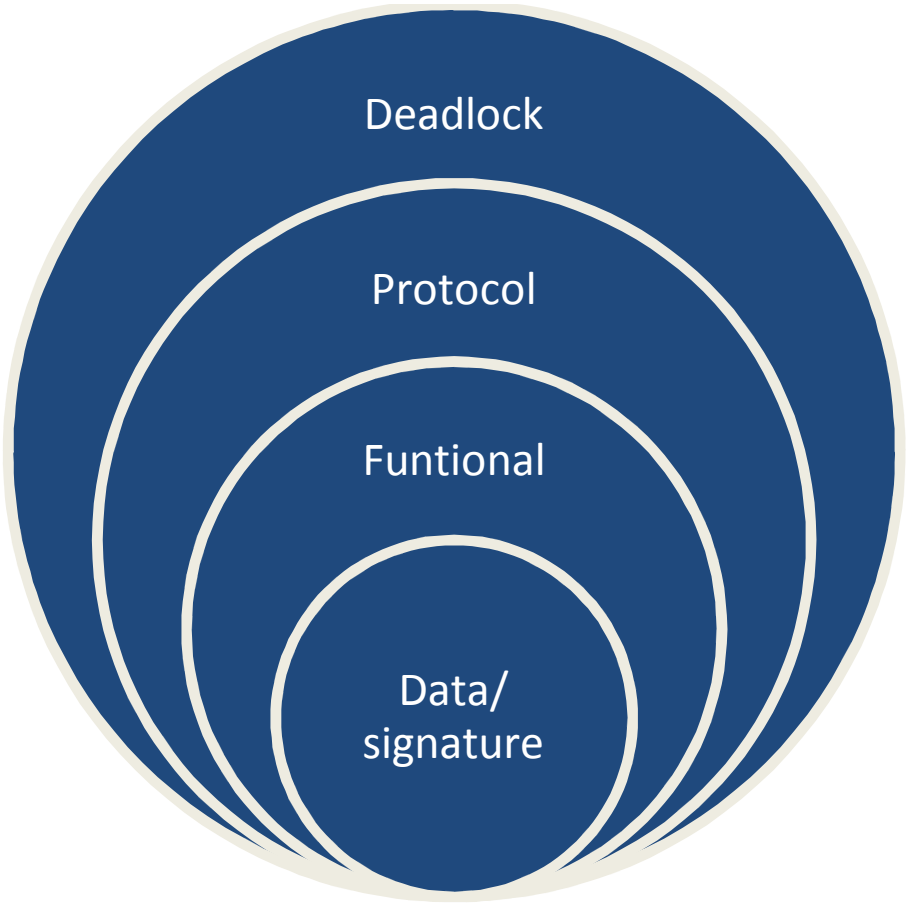

Figure 1. Hierarchy of mismatches in behavioral analysis

Based on the provided definitions, it shows that all four levels of mismatches are needed to be analyzed and rectified in order to support the behavioral compatibility between web services. Figure 1 shows the hierarchy of the levels of mismatches to support behavioral analysis. It illustrates that the signature/data level of mismatches in a web service is the solution for resolving behavioral analysis. This level needs to be first analyzed in order to resolve the other levels of mismatches. It is then followed by the functional mismatch which is important to the meaning of the messages or the overall task of web services. The protocol level mismatches can be identified and resolved correctly by giving importance to the signature and semantic level mismatches. Finally, the number of deadlock mismatches can be reduced by focusing on the inner levels of the mismatches which contribute to the deadlock situations such as the data, functional and protocol level of mismatches.

## 2.3 Resolving Behavioral Incompatibility

Once the causes of behavioral incompatibility in the web service environment have been identified the next step to take would be to address this incompatibility. This section describes how mismatches are addressed and resolved in order to facilitate successful communication between the service requestor and the provider. There are two terms that have been widely adopted and they are the adaptation and mediation which are steps to resolve these behavioral mismatches.

### 2.3.1 Adaptation

The concept of adaption has been widely discussed in the system integration initiatives in CBSE. There are many approaches that have used the adaptation concepts in web services of [21], [14],[22], [23] and [24]. According to[11], service adaptation refers to *generating a new service (adaptor) that supports interoperability among web services with different interfaces and protocols during the service interaction*. [14] refer to software adaptation as *a promising solution to compose web services with interface mismatches by generating a new component called adaptor*. They also claim that the generated adaptor acts as an orchestrator to make the involved services work correctly.

### 2.3.2 Mediation

The concepts of mediation in handling and resolving mismatches and heterogeneity problem was introduced in early 1990 by Wiederhold in database management systems[25]. According to Wiederhold "*mediation makes an interface intelligent by dealing with representation and abstraction problems that you must face when trying to use today's data and knowledge resources*" and he has defined mediator as a *software module that exploits encoded knowledge about certain sets or subsets of data to create information for a high layer of application*[25].

The mediation terminology has been reintroduced in web services by [6]. They have described mediation as "*process for settling a dispute between two parties where a third one is employed whose task is try to find common ground that will resolve inconsistencies between their respective conceptualizations of a given domain*". Apart from the definitions of Fensel and Bussler, there are many other definitions for mediator in the context of web services. For instances, [26]define mediator as *software module that provides sharing of services and agglomeration of resources into complex service* and [5] define mediation *as ability to resolve heterogeneities among heterogeneous entities which allow parties to exchange messages, documents and data disrespect of their vocabulary and behavioral models*. There are many other web service approaches that have called the solution for behavioral mismatches as mediator, such as [18, 27-30].

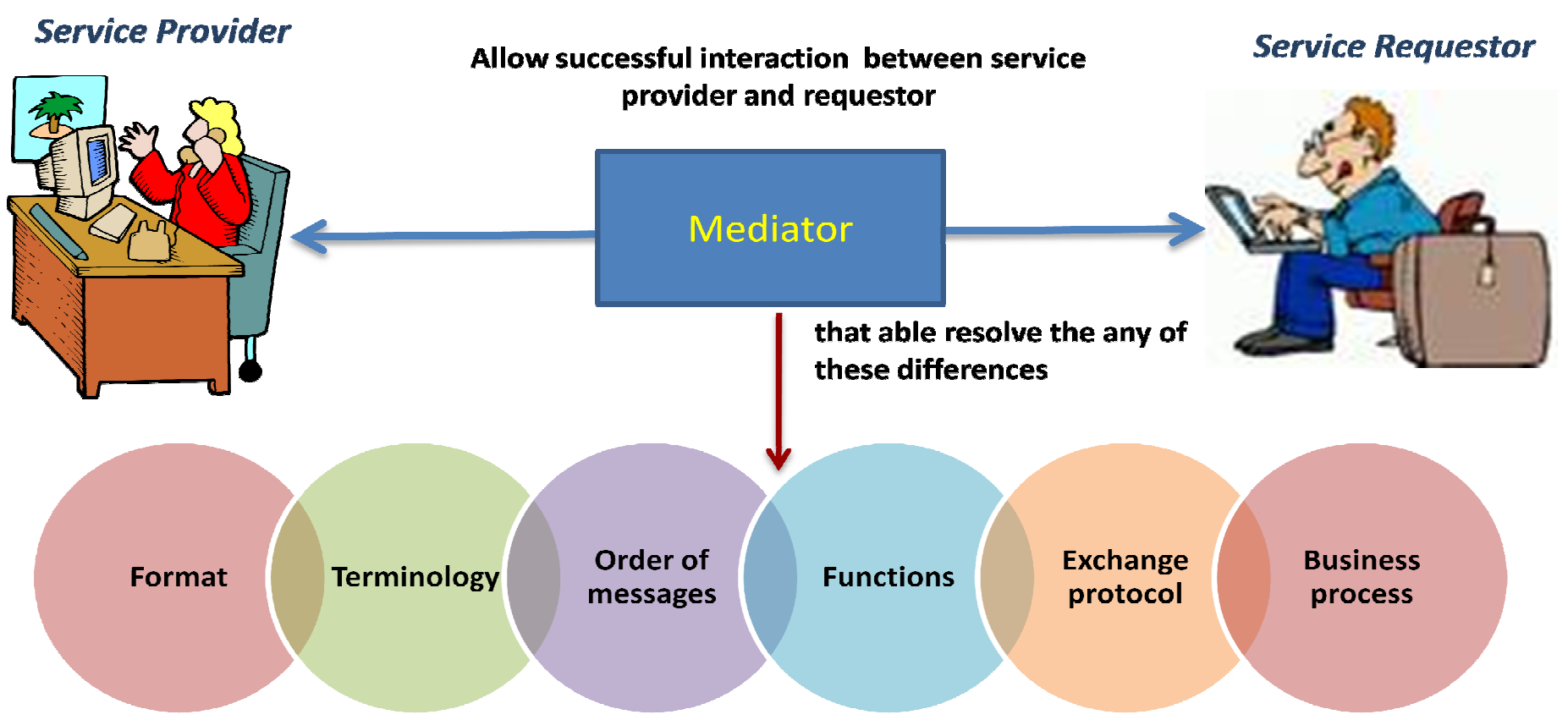

Figure 2. Overview of Mediator in Web Services

Based on the provided definitions for both adaptation and mediation, it can be concluded that these definitions have the same objective which is in resolving the incompatibility between the interacting web services to provide successful communication. Figure 2 illustrates the role of the mediator in web services.

Therefore, this paper includes both the adaptation and mediation approaches that support behavioral mismatches in the context of web services. Here on, both these terns will be referred to as "mediator" or "mediation" while the whole process of identifying and resolving the web service behavioral incompatibility will be referred to as "process mediation."

## 3. SYSTEMATIC REVIEW OF PROCESS MEDIATION APPROACHES

This section will present the state-of-art approaches that produce solutions for the process mediation and some early academic based research works that are useful in understanding the process mediation concepts and features. Early research in identifying behavioral incompatibility has been introduced by [7] for generating software adapters to bridge different software components. However, [7] assume that the messages uses the same parameter during the generation of the adaptor. [31] has classified behavioral mismatches into five different patterns on their respective adapter templates and claimed that automation in web service compatibility analysis, involves complex processes. Similarly, [32] have also identified the specific type of behavioral mismatches and proposed a general algorithm to resolve them in the WSMO Framework. Another early process mediation approach was introduced by [15]. This approach proposed six operators to provide mapping between the mismatched interacting services. Then it automatically manages the messages based on the logics in the transformation expressions by storing, transforming and forwarding the messages accordingly between the web services.

Apart from these approaches, twelve state-of-the-art approaches that support process mediation in web services using System Literature Review (SLR) technique have been carefully selected. The objective of this systematic review is to identify the state of art process mediation approaches in web service or the semantic web service. Two important research questions have arisen and they are as follows.

- What are the published process mediation approaches for the semantic web service?
- What are the methods or techniques involved in the semantic web service process mediation?

It is believed by the authors of this paper that the effort and time spent to answer these questions on this systematic review can lead them towards reaching the objective of their study. Their search of the sources mainly focuses on the electronic databases such as ACM, IEEE Explore, Springer Link, Web of Science and Science Direct using SCOPUS electronic database. These have been selected as they are in the most relevant journal articles, lecture notes and conference papers from 2000 to 2010 and are included in this systematic review as the latest literature. This research is limited to the area of computer science, though mediation is also one of the important keywords in social science. A few selected keywords that are most relevant in studying such terms as "mediation", "adaptation", "compatibility" and "interoperability" which append with "web services" and "semantic web services". The keyword search has resulted in 1860 papers and only twelve papers are selected as the state-of-art approaches in the comparative evaluation carried out for this paper. Table 2, shows how the 1860 papers were narrowed down to only twelve papers.

Table 2. Search results

| **Total papers** | **Number of paper** |
|---|---|
| Search by Keyword | 1860 |
| After removing irrelevant titles | 337 |
| After removing not relevant abstract | 172 |
| After browsing through the full paper | 110 |
| Selected papers | 12 |

The focus basically is on process mediation in the software systems based on web service. The papers that cater for adaptation or mediation for multimedia, mobile environment, Geographic Information Systems (GIS), ReSTful System and network protocol have been removed and the papers that discuss context-based and timed protocol solution in resolving behavioral compatibility have also been omitted. This systematic review of techniques has also helped in identifying all the related resources to support the comparative evaluation conducted. Figure 3 shows the selected twelve state-of-art approaches that are categorized as the year of publication from 2007 to 2010. A unique number is given for each paper to ease the comparative evaluation in Section 4.

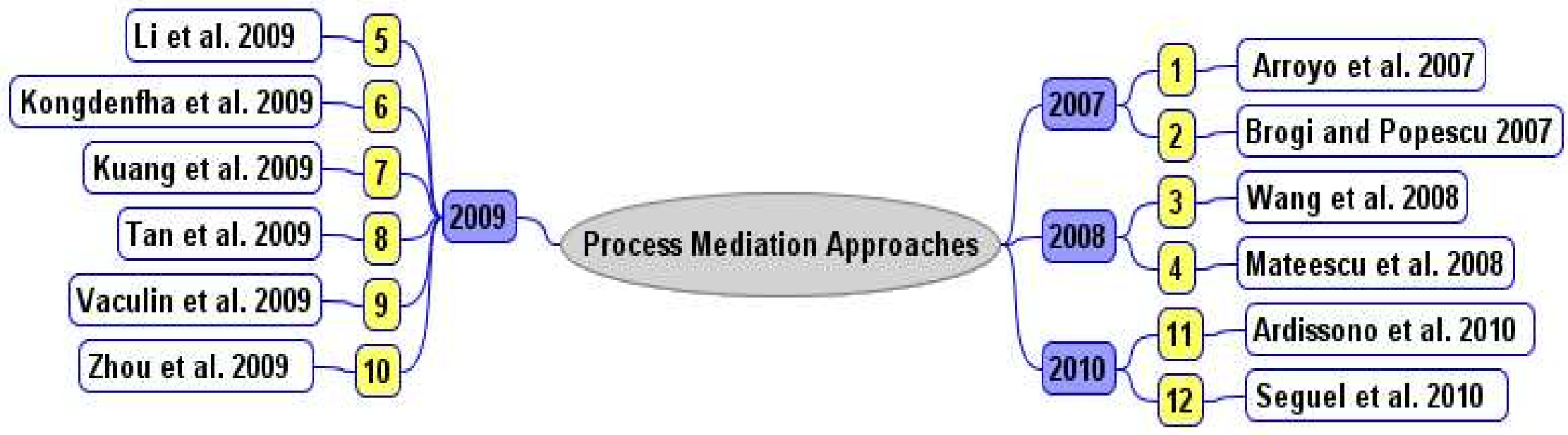

Figure 3. Selected state-of-art process mediation approaches

## 4. OVERVIEW OF PROCESS MEDIATOR

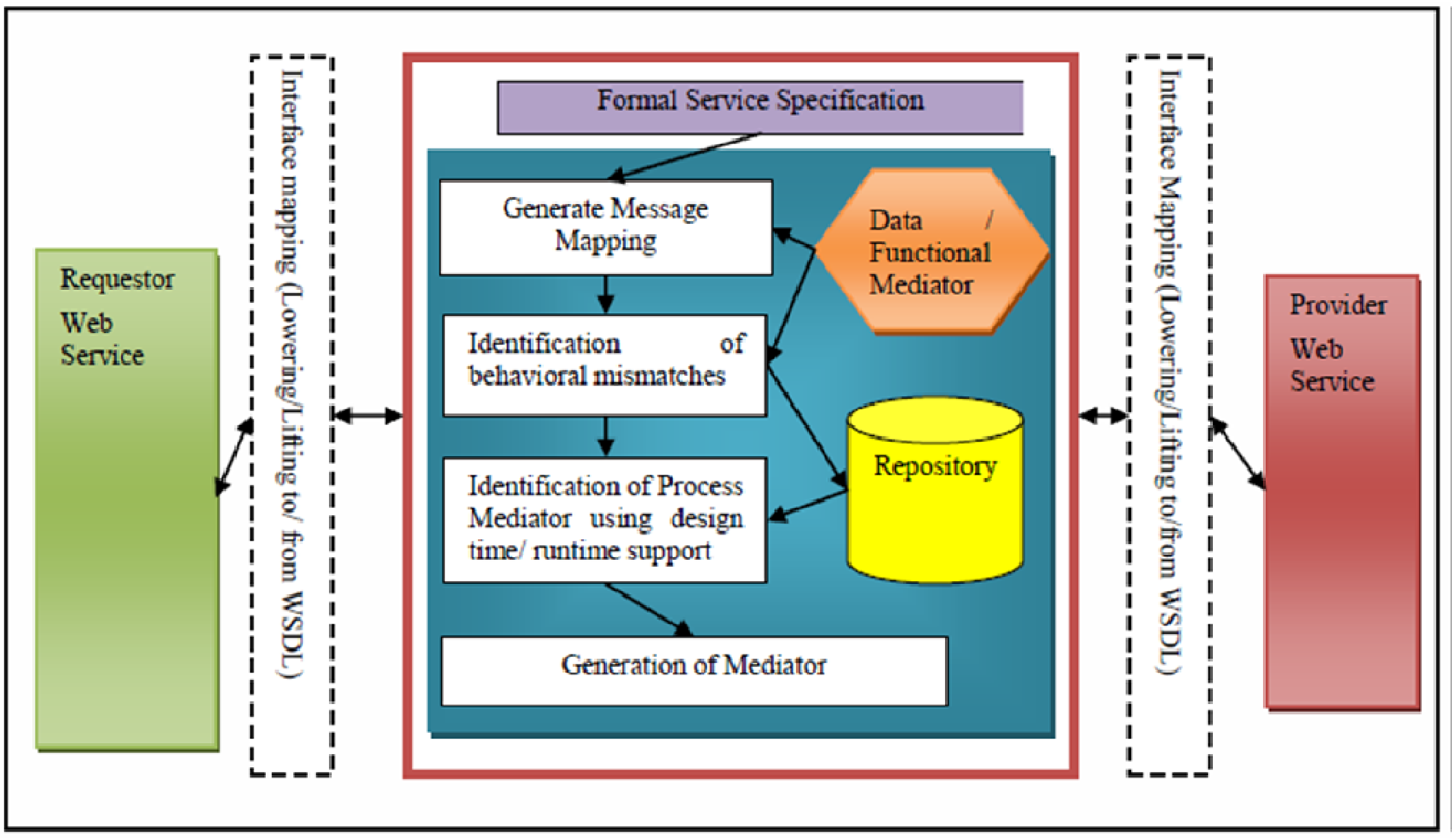

Figure 4. Overview of Process Mediation

This section provides an overview of the process mediator in the web service. A few important process mediation components have been identified based on the analysis of the state-of-art process mediation approaches. Figure 4 illustrates the overview of the process mediation and each component is explained below. There are seven important components that have been traced and these are: the formal service specification, the data/functional mediator, the repository, generating message mapping, the identification of mismatches, and the identification of process mediator and the generation of mediator. This section explains the first three components and the rest of the components will be explained in detail in Section 5.

### 4.1 Formal Service Specification

Generally, most of the approaches, transform the web services into formal specifications using formal models. These formal specifications are widely used due to the strong mathematical theorems behind them. The formal model is able to facilitate process incompatibility analysis and reduces ambiguity during the generation of the process mediator. There are three famous categories of the formal models that can be used in formalizing the process interaction namely; the Automata, the Process Algebra and the Petri Net.

There are three well known automata based methods which are the Abstract State Machine (ASM), the Labeled Transition System (LTS) and the Finite State Machine (FSM). Generally, all the automata based methods represent the building blocks of the web service process in a simple graphical format by illustrating the data flow, control flow and transition between states or action. ASM has been adopted as a basic mechanism for modeling the interaction between the web services in process mediation solutions. This mechanism is successfully used in the WSMO based approaches.FSM has been used by [33],[22] and LTS has been applied by [14] and [18, 30] [27] have used Petri Net based formalizations called Coloured Petri Net (CPN) and Service Net to analyze behavior incompatiblity between web services during the web service composition. There are few tools which have been identified and are that able to transform the web services into formal models such as BPELtoCPN and BPELtoSTS transformer [14].

### 4.2 Data / Functional Mediator

Most of the process mediation approaches do not provide any explicit description of the data or the functional mediator due to two reasons. Firstly, it has been argued that generating data and the functional mediator can both be huge research topics on their own. Secondly, some approach gurus are under the assumption that the web services have compatible messages at the data and functional levels. Therefore process mediation is only discussed from the protocol level mismatches. Some approaches have adopted the data / functional component as external components or agents to support the protocol and deadlock mismatches. As discussed earlier there a few available tools that support data mediators such as Microsoft Biztalk and Stylus Studio XML Schema Mapping. However, we do not come across any tools that support the functional mediator. Both the data and functional mediators play important roles in generating message mapping and identifying the correct behavioral mismatches.

### 4.3 Repository

The repository component plays an important role in process mediation by storing message elements, message mapping, mapping rules, mismatch patterns, data flow and the control flow patterns between the web services. Most of the approaches solve the ordering and missing message mismatches by storing the messages in the repository and retrieving them as needed. In addition, [31] have used information called evidence that is stored in the repository to resolve deadlocks.

## 5. COMPARATIVE EVALUATION OF PROCESS MEDIATOR APPROACHES

This section provides comparative evaluation of process mediation approaches that have been selected in Section 3. These approaches are evaluated based on the evaluation criteria of four important components of process mediation; namely message mapping, identification of protocol mismatches, identification of the process mediator and the generation of the process mediator. Figure 5 illustrates the evaluation criteria of each of the four process mediation components that were mentioned earlier.

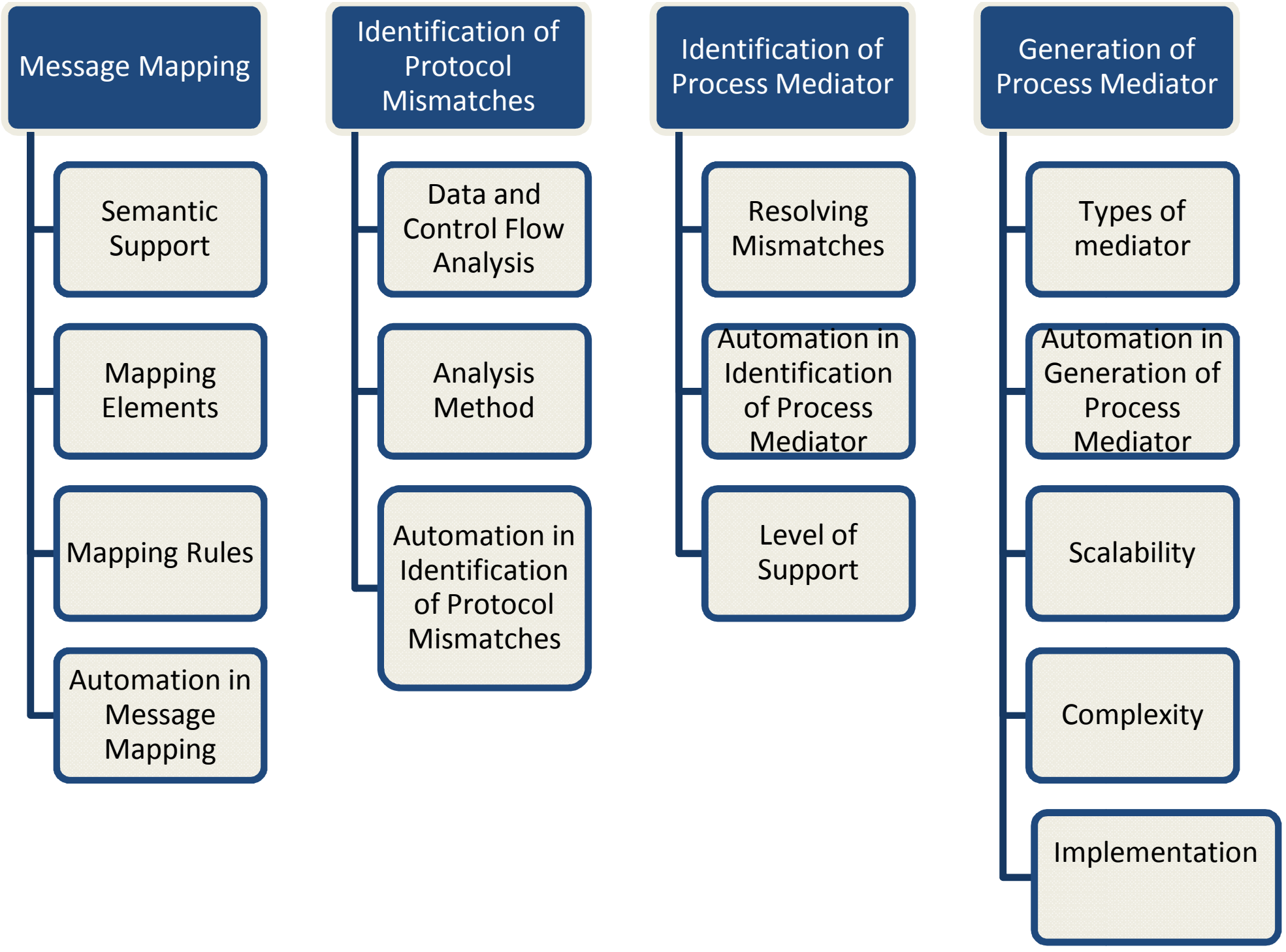

Figure 5. Evaluation criteria for each component

## 5.1 Message Mapping

Message matching is an important step in producing a process mediator. Behavioral compatibility between web services can be identified by the matching of messages that are involved in the interaction. Message matching involves data mapping (data mediation) and functional mapping/mediation in order to determine the similarity or differences in the operation name and input/output elements used in the web services. Generally the data and functional mapping only discusses implicitly in the process mediation approaches. There are two approaches one by [30] and the other by [24] where both do not mention message mapping since they only focus on protocol mismatches by assuming that there is no incompatibility in the data or the functional levels. Table 3 provides the comparative evaluation for the message mapping component. Below are a few elements that are taken into consideration in order to provide the message mapping between the web services:

a) semantic/syntactic support in order to generate message mapping
b) elements that involved in the message mapping
c) rules that apply in message mapping
d) automation level in message mapping

### 5.1.1 Semantic Support

Semantic support is essential in order to provide automation in message mapping and improve the correctness of the mapping provided. As discussed earlier, most of the approaches assume that the operation and the parameter have identical names. Only three approaches namely [28], [21], [23] and [5] have provided ontology based on the semantic support in producing message mapping.

### 5.1.2 Mapping Elements

There are two types of mapping elements that have been widely used in the process mediation approaches as listed below.

- Input / output –input and output of the related web service operation are directly used in order to identify the mismatches without any transformation.
- Input, Output, Precondition and Effect (IOPE) - the basic input and output of the operations are transformed into IOPE before identifying the message mapping. Generally, these approaches have adopted the automata based formal modeling; added the precondition and effects according to the messages that contribute to a particular state.

### 5.1.3 Mapping Rules

There are only six approaches that have explained the rules that have been applied in identifying the message mapping in the web service application. Below are the rules that have been explained by these approaches:-

- output elements of a source service matched with the input elements of the target web service as explained by [5], [11], [23], [27]and [29].
- [28] has three rules in generating message mapping. Firstly, all the input requested from the provider needs to be delivered by the requestor; secondly, the preconditions of a state need to satisfy the requestor and finally all the output and effects that are requested by the requestor need to be provided by the provider.

### 5.1.4 Automation in Message Mapping

Most of the message mapping in the process mediator is provided by the developers who have sufficient knowledge on the domain of the specific scenario. The developers provide a mapping table for each operation in the sourced web service by the corresponding operation in the targeted web service. Apart from the manual identification of message mapping, [11] have used the scoring method based on the scheme mapping, [28] have used rules in order to find the message mapping while [5] and [21] have used reasoning based on the ontology mapping in order to provide some level of automation in producing the message mapping.

Table 3. Evaluation based on message mapping criteria

| Main Features | Evaluation Criteria | Approaches | | | | | | | | | | | |
|---|---|---|---|---|---|---|---|---|---|---|---|---|---|
| | | 1 | 2 | 3 | 4 | 5 | 6 | 7 | 8 | 9 | 10 | 11 | 12 |
| Automation in the generation of message mapping | By Developer | | √ | | √ | √ | | √ | √ | | n/a | √ | n/a |
| | Automation (semi) | √ | | √ | | | √ | | | √ | n/a | | n/a |
| Semantic Support | Syntactic | | √ | | √ | √ | √ | | √ | | √ | √ | √ |
| | Semantic | √ | | √ | | | | √ | | √ | | | |
| Mapping Elements | Input and Output | √ | √ | √ | √ | √ | √ | √ | √ | √ | n/a | √ | n/a |
| | Precondition and Effect | | √ | | | | | | | √ | n/a | | n/a |
| Mapping Rules | Provided Rules | √ | n/a | n/a | n/a | n/a | √ | √ | √ | √ | n/a | √ | n/a |

## 5.2 Identification of Protocol Mismatches

This section presents the method of identifying the protocol mismatches. Table 4 provides the comparative evaluation for the identification of protocol mismatch components. There are three criteria that have been analyzed in order to identify the protocol mismatches in web services as listed below namely, data and control flow analysis, the analysis method and the automation in identification of the protocol mismatches.

### 5.2.1 Data and Control Flow Analysis

Data flow analysis helps the behavioral compatibility study based on the message exchanged between the web services. The messages exchanged between the provider and requestor of the web service is also identified as the operation that sends and receives between the target and the sourced web service. The control flow analysis ensures the process mediator supports the basic business logic in the web service interaction. The control flow analysis provides logical flow business process of each web service by addressing the flow constraints such as ***sequence*** (for serial execution), ***while*** (to implement a loop), ***switch*** (for multiple way branching), ***flow*** (for parallel execution) and ***pick*** (for choosing among the alternative path). Figure 6 and 7 illustrate the data and control flow analysis in the web service environment. Data flow and control analysis in process mediation is provided by basing on the visibility of the web service as discussed by [28]. The web services with limited visibility only provide the exchanged operations to allow data flow analysis during the generation of a process mediator. Thus, the web services with complete visibility allow behavioral analysis in order to generate process mediator based on both data and control analysis perspectives.

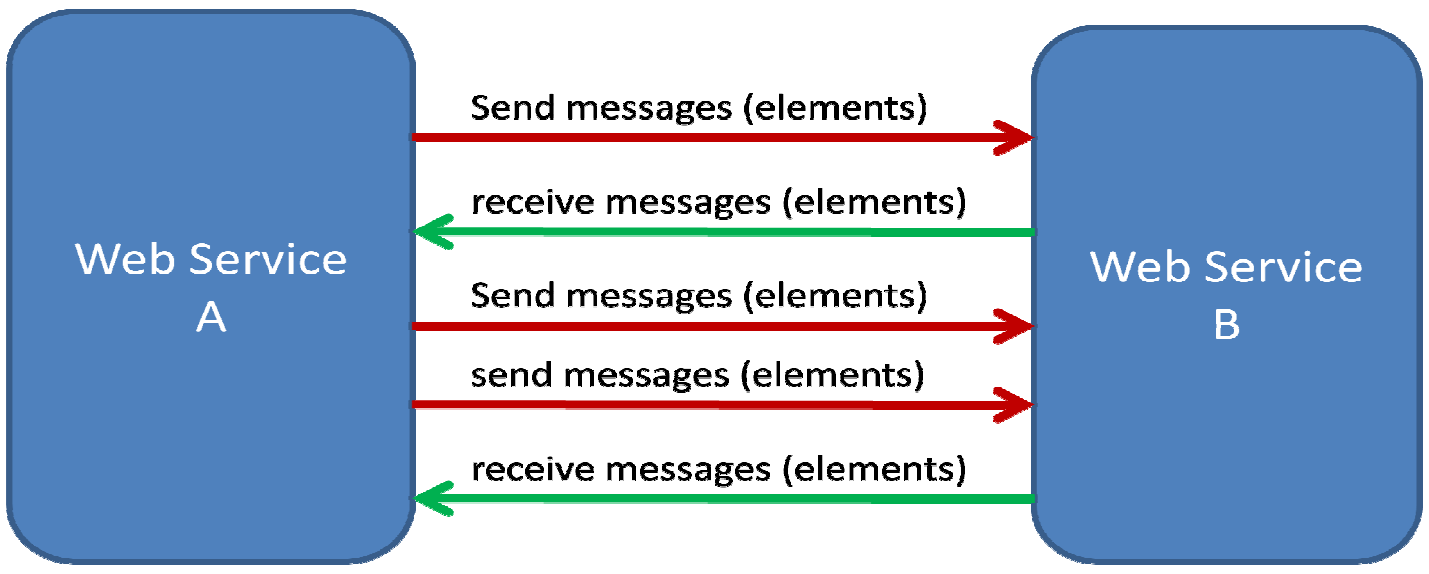

Figure 6. Data flow analysis in web service interaction

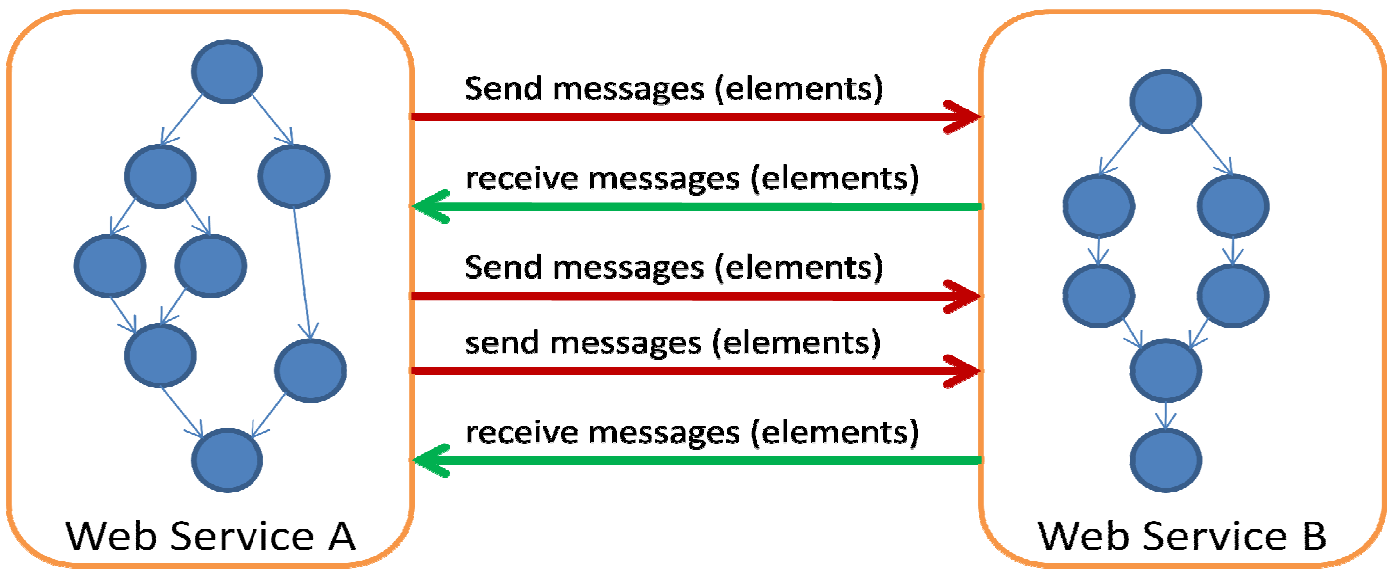

Figure 7. Data and Control flow analysis in web service interaction

### 5.2.2 Analysis Method

There are two main methods that have been adopted in order to identify the process mediator in the web service interaction namely the Specific Protocol Analysis (SPA) and the General Reachability Analysis (GRA).

### 5.2.2.1 Specific Protocol Analysis (SPA)

The approaches that use SPA method identifies each protocol mismatch that have been explained individually in Table 1. As the result, produced specific patterns have been produced in order to identify the mismatches and to produce the resolutions for each of them. Table 4 shows the process mediator patterns that are proposed by these approaches. These patterns have been identified using a developer's decision [34], tokens [29], adapter template [22] and logic boxes [5] based on the message mapping on the particular application domain. In addition, [23] has also used this method without providing any specific name for the identified mismatches. Similarly, [22] also do not provide any specific name for the wrong order mismatches, but has provided a solution for this mismatches by storing the input from the source / target messages and directing them to the right recipient.

Table 4. Various patterns that supports protocol mismatches

| No | Protocol Mismatches | [5] | [34] | [22] | [27] | [29] |
|---|---|---|---|---|---|---|
| 1 | Extra Messages (EM) | N/A | Simple StorerPattern (SSP) and SCP | Extra Message Pattern | N/A | Zero-to-one mapping |
| 2 | Missing Messages (MM) | Add Box | Simple Constructor Pattern (SCP) and SSP | Missing Message Pattern | N/A | One-to-zero mapping |
| 3 | One-to-Many Messages (OMM) | Split Box | Splitter Pattern | One-to-Many Pattern | Split Mediator | One-to-N mapping |
| 4 | Many-to-One Message (MOM) | Merge Box | Merger Pattern | Many-to-One Pattern | Merge Mediator | N-to-one mapping |
| 5 | Wrong-Order-Messages (WOM) | Select Box | Simple Storer Pattern | Ordering Pattern | Transformation Mediator | N/A |

### 5.2.2.2 General Reachability Analysis (GRA)

The second method in identifying suitable process mediation is based on the General Reachability Analysis (GRA). The approaches that adapt this method first determine whether the overall web service interaction is successful or not by ensuring there is no deadlock communications or information loss is exist in the interactions. The main objective is to provide a successful communication path by identifying and preventing unsolvable web service interaction. Therefore, these approaches pay more importance to finding a reachable communication path than identifying the specific protocol mismatches. Directed graphs and search space technique are used to explore all the executable paths in the web service execution.

- **Directed Graphs** – [21] has used a Reachability Graph (RG) which is derived from the YAWL workflow to provide a visual aid in identifying the suitable mediator and the required interaction. [27] uses the concept of Communication Reachability Graph (CRG) to ensure the existence of a process mediator before constructing mediator. [30] and [24] use Dependency Graphs (DG) to eliminate irrelevant interactions and construct a path that satisfies all the dependencies.

- **Search Space -** [33] has used this technique to traverse a list of messages, and has also provided backward-chaining algorithm to identify deadlock and information loss.[35]used this technique to identify traces that represent the correct interaction between web services. [28] uses search space to find the possible mapping in the requestor's execution path.

There are two issues that are related to the GRA method. Firstly, the GRA method is useful in any phase of process mediation. It can be used to identify the executable paths that have mismatches and disallow the unexpected output from being generated. It can also be used after the generation of the process mediator to ensure correctness. Most of the approaches in this category use some additional rules to identify and resolve the specific protocol mismatches. Only [21] and [24] do not discuss any individual protocol mismatch identification. Secondly, the approaches that use the GRA method need to provide a mechanism to handle complexity. This is due to the complexity of the Directed Graph and the Search Space Algorithm that increase the exponents according to the number of nodes or state web service interaction. Thus, these approaches need to provide a complexity reduction technique to support complex and real world web service interactions.

### 5.2.3 Automation in Identification of Protocol Mismatches

This section provides an analysis on how these process mediators are being identified according to specific protocol mismatches. Generally these process mediators are identified manually by the developer or automatically by using logic based algorithms or software tools. [14], [33], and [18] are fully dependent on developers to identify suitable mediators according to the protocol mismatches.

Table 5. Evaluation based on identification of protocol mismatches criteria

| Main Features | Evaluation Criteria | Approaches | | | | | | | | | | | |
|---|---|---|---|---|---|---|---|---|---|---|---|---|---|
| | | 1 | 2 | 3 | 4 | 5 | 6 | 7 | 8 | 9 | 10 | 11 | 12 |
| Support Flow Analysis | Data flow | √ | √ | √ | √ | √ | √ | √ | √ | √ | √ | √ | √ |
| | Control Flow | | √ | | √ | √ | | | √ | √ | √ | √ | √ |
| Analysis method | Specific Protocol Analysis | √ | | | | √ | √ | √ | √ | | | √ | |
| | General Reachability Analysis | | √ | √ | √ | | | | √ | √ | √ | | √ |
| Automation in mismatch identification | Identified by developer | | √ | √ | √ | √ | | | | | | | |
| | Semi-automatic | √ | | | | | √ | √ | √ | √ | √ | √ | √ |

## 5.3 Identifying the Process Mediator

This section presents the methods of identifying process mediators for a specific protocol mismatch. Table 6 provides the comparative evaluation for identification of the process mediator. There are three criteria which have been analyzed in order to the identify protocol mismatches in web services namely, resolving protocol mismatches, automation in identifying a process mediator and the level of support.

Table 6. Evaluation based on identification of the process mediator criteria

| **Main Features** | **Evaluation Criteria** | **Approaches** | | | | | | | | | | | |
|---|---|---|---|---|---|---|---|---|---|---|---|---|---|
| | | **1** | **2** | **3** | **4** | **5** | **6** | **7** | **8** | **9** | **10** | **11** | **12** |
| Resolve protocol mismatches | Extra Messages (EM) | × | **n/a** | √ | × | √ | √ | √ | √ | √ | √ | √ | **n/a** |
| | Missing Messages (MM) | √ | **n/a** | × | × | √ | √ | × | × | × | √ | √ | **n/a** |
| | Many-to-One Messages (MOM) | √ | **n/a** | √ | × | √ | √ | √ | √ | √ | × | √ | **n/a** |
| | One-to-Many Messages (OMM) | √ | **n/a** | √ | × | √ | √ | √ | √ | √ | × | √ | **n/a** |
| | Wrong Order Messages (WOM) | √ | **n/a** | √ | √ | √ | √ | √ | √ | √ | √ | √ | **n/a** |
| Automation in protocol mediator identification | Identified by developer | | | √ | √ | √ | | | | | | | |
| | Automatic (Semi) | √ | √ | | | | √ | √ | √ | √ | √ | √ | √ |
| Level of Support | Design time | √ | √ | | √ | √ | | | | | √ | | √ |
| | Run time | √ | | √ | | | √ | √ | √ | √ | √ | √ | √ |

### 5.3.1 Resolving Mismatches

This section provides general techniques that have been widely adopted by approaches in order to address each of the protocol mismatches. Generally EM and WOM have been addressed by storing and forwarding the incoming and outgoing messages. Most approaches provide additional repository to store the incoming and outgoing messages. These stored messages are only forwarded to the needed web services to handle the wrong order mismatch or an extra message. The MOM and OMM mismatches are highly dependent on the determined message mapping. It is a difficult task for a web service to merge and split the messages automatically and correctly at runtime to support successful interaction. Similar to the EM and WOM, MM can also be resolved by using the storing and forwarding techniques but only when the missing messages are available in the repository. Again, it is a hard task for a web service to automatically generate a new web service message with the correct elements without the developer's support at the time of design. The available approaches like [22] and [27] were able to produce acknowledgement based messages with the support of the stored input elements.

### 5.3.2 Automation in the Identification of the Process Mediator and the Level of Support

This section provides an analysis on how the process mediators are being identified according to their specific protocol mismatches. Generally these process mediators are identified manually by the developer or automatically using the logic based algorithms or software tools. [14], [33], and [18] fully depend on the developer to identify the suitable mediator according to the protocol mismatches. Approaches from [5], [22], [23], [27], [28], [29] and [24] use logic based algorithms to identify the solution for the interaction mismatches. On the other hand, [21] and [30] provide automatic solution through software tools which provide support at design time.

## 5.4 Generation of Process Mediator

This section provides a few important features of the process mediator approaches such as the types of the produced mediator, automation in producing the mediator, scalability, complexity and the actual implementation of the approaches as listed in Table 7. Generally, the generated mediator can be in two different forms for the identified mismatches as stated below.

Table 7. Evaluation based on the generation of the process mediator

| Main Features | Evaluation Criteria | Approaches | | | | | | | | | | | |
|---|---|---|---|---|---|---|---|---|---|---|---|---|---|
| | | 1 | 2 | 3 | 4 | 5 | 6 | 7 | 8 | 9 | 10 | 11 | 12 |
| Type of Mediator | New service / contract / specification | **n/a** | | √ | √ | √ | √ | √ | √ | √ | √ | √ | √ |
| | Modifying existing service | **n/a** | √ | | | | | | | | | | |
| Automation in process mediator generation | Generated by tools | | √ | | √ | √ | √ | | √ | √ | | √ | √ |
| | Not specified | √ | | √ | | | | √ | | | √ | | |
| Scalability | Specific Scenario | | √ | √ | √ | √ | √ | √ | √ | √ | √ | √ | √ |
| | Generic Scenario | √ | | | | | | | | | | | |
| Complexity | Small case study | √ | √ | √ | √ | | √ | √ | √ | √ | √ | | √ |
| | Large case study | | | | | √ | | | | | | √ | |
| Implementation | Prototype | √ | √ | √ | √ | | | √ | | √ | √ | | |
| | Implemented with evaluation | | | | | √ | √ | | √ | | | √ | √ |

Firstly, the process mediator can be a new service that mediates the interaction between incompatible web services. This method is able to support needed changes in each service without disrupting other faultless interactions. However, it adds the maintainability issues which are needed to be tackled when the original web service evolves. Almost all the state-of-art approaches have adopted this method in order to generate a process mediator.

The second form of process mediator is generated by modifying business logic of one service to make it compatible with other services. However, the second method is not widely adopted as compared to the first method since any small change in a service may cause a significant impact on potentially thousands of other services which are not prepared for the change. Therefore, implementation of this method is expensive and error-prone due to the versioning of the business logic to cater for all the interacting web services. This method is acceptable when a service only interacts with one selected service. This method have been applied by[22].

Most of the approaches provide automatic generation of the process mediator based on the identified patterns, mediator specifications and contracts using software tools. There are four approaches that do not provide any details on how the process mediator is being generated namely by [5], [33], [23] and [30]. The scalability of the selected approaches have been evaluated and the criteria based on the type of scenario, the proposed process mediator's ability to handle; whether any solutions could be provided for any generic scenarios or for any specific scenario was also analyzed.. Based on this evaluation, only [5] were able to provide solutions for the generic scenario and the rest of the approaches focused on resolving specific scenarios such as the purchase request, online traveling and shopping. This means that the approaches that only support the specific scenario need to be studied and implemented all over again to support the different behavioral mismatches in other web service scenarios.

The next step taken was to evaluate the complexity based on the size of the case study that the researchers' had chosen to implement the process mediator. No specific scale was used to classify the size of the case study into small or large classification. The criteria was evaluated basing on the observation of the number of nodes, number of web services involved and the authors' comment in the conclusion and future work. Based on this analysis only two approaches have been identified i.e. [18, 29] are able support the behavioral mismatches in complex web services.

Finally, the actual implementation of the approaches was evaluated. It was found that only 5 approaches have fully been implemented and compared with other process mediation approaches. The rest of the approaches were only implemented as prototypes.

## 6. Discussion

This section provides a summary on the comparative evaluation which is presented in Section 5. Eight important elements in a process mediator that support behavioral mismatches in web services with/without a minimum involvement of the developer have been chosen. These eight elements and the comparative study are listed in Table 8. The automation criteria have been divided into four phases namely, the automation in generating message mapping, identifying the mismatches, identifying the process mediator and finally generating the process mediator. The comparative evaluation based on the support for control and data flow, semantic and reasoning, generic scenario, complex case study and lastly on the evaluated implementation has also been provided.

Table 8. Summary of the comparative evaluation

| Discussion elements | Approaches | | | | | | | | | | | |
|---|---|---|---|---|---|---|---|---|---|---|---|---|
| | 1 | 2 | 3 | 4 | 5 | 6 | 7 | 8 | 9 | 10 | 11 | 12 |
| Support for both control and data flow | × | √ | × | √ | √ | × | × | √ | √ | √ | √ | √ |
| Support for semantic and reasoning | √ | × | √ | × | × | × | √ | × | √ | × | × | × |
| Automation in the generation of message mapping | √ | × | √ | × | × | × | × | × | √ | × | × | × |
| Automation in mismatch identification | √ | × | × | × | × | √ | √ | √ | √ | √ | √ | √ |
| Automation in the identification of the process mediator | √ | √ | × | × | × | √ | √ | √ | √ | × | √ | × |
| Automation in generating the process mediator | × | √ | × | √ | √ | √ | × | √ | √ | × | √ | √ |
| Scalable to support generic scenario | √ | × | × | × | × | × | × | × | × | × | × | × |
| Support complex case study | × | × | × | × | √ | × | × | × | × | × | √ | × |
| Implemented with evaluation | × | × | × | × | √ | √ | × | √ | √ | × | √ | √ |

Generally, the evaluation shows that none of the approaches cover all the evaluation criteria. [28, 29] were able to support most of the process mediation elements and fully implemented them with an evaluation. [5] has proposed a high level of automation in their approach. However their approach neither provides explicit explanation on their implementation of their approach nor presents sufficient evaluation. There are a few important factors that have been identified in this evaluation and they are:

- the use of semantics play an important role in improving the automation level in the process mediation task in web services.
- the generation of message mapping is essential to support other automation tasks in process mediation but however the generation of message mapping is widely identified by developers rather than the one that is automatically generated

In this comparative evaluation, no measurable evaluation as high or low is provided, since the objective of this review is to identify areas in process mediation in web services that are under explored. A gap analysis has been formulated for the process mediation by using a spider graph chart as shown in Figure 9. Based on this spider graph, it can be concluded that there are four important areas in process mediation in web services that are not fully explored. There are namely, the automation in message mapping generation, the scalability to support generic scenario, the support for semantics and reasoning and finally the support for complex web service case studies.

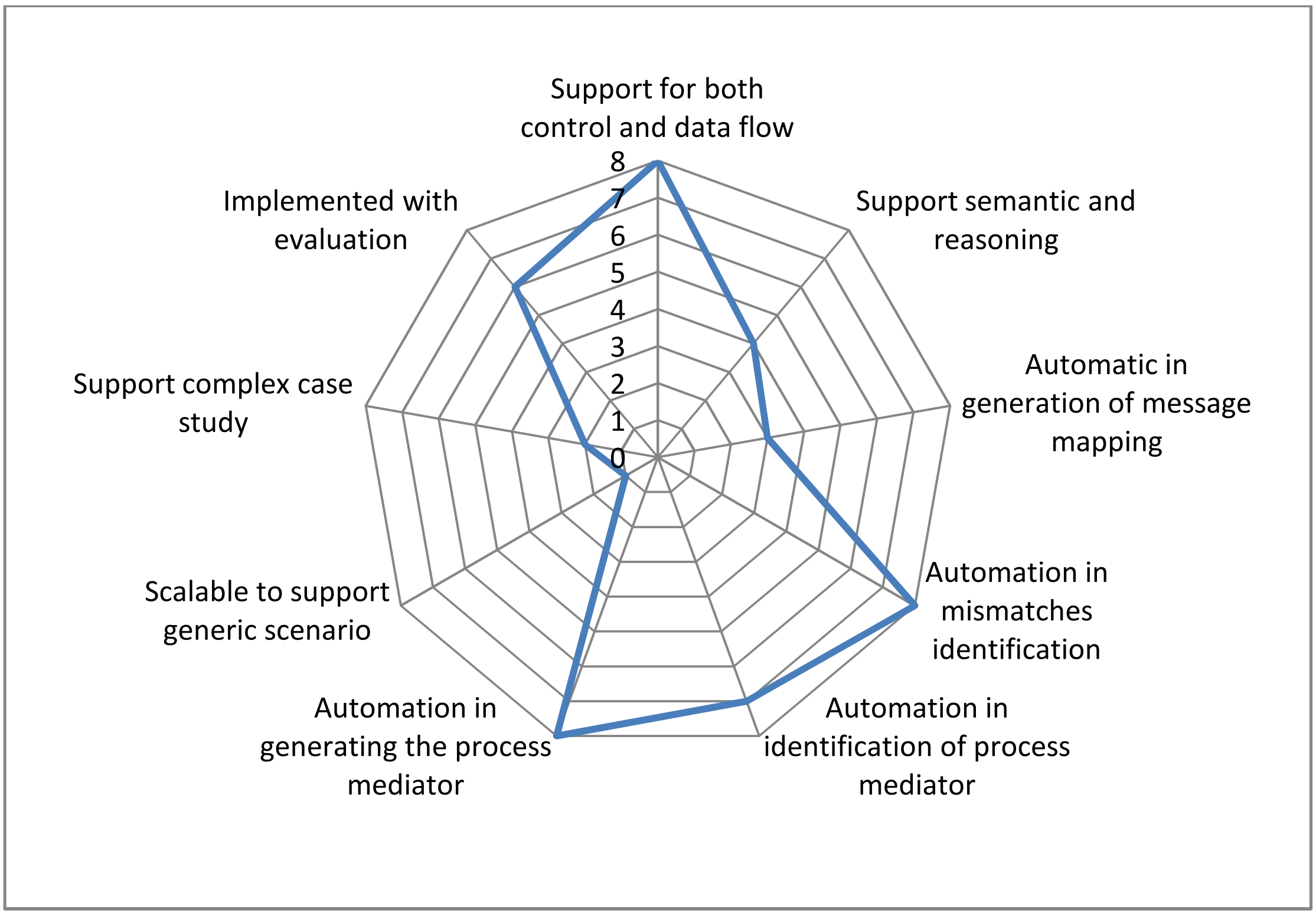

Figure 9. Gap analysis using spider graph

## 7. Conclusions

The main objective of this paper is to conduct a comparative study on the state-of-art process mediation approaches and provide a gap analysis. In this paper, firstly the behavioral incompatibility in the context of web services has been defined. Then the selection of process mediation approaches based on SLR technique is presented. Then, an overview for process mediation by describing the four important components of process mediation such as message mapping, identification of protocol mismatches, identification and generation of process mediation have been provided. In addition, all the current approaches based on some process mediation components as explained earlier have been analyzed. Finally, the comparative evaluation is summarized by providing a gap analysis that identifies the process mediation areas that are not fully explored.

## Acknowledgements

This research is supported by Fundamental Research Grant Scheme (FRGS) [grant number 4F054] from the Ministry of Science and Technology and Innovation (MOSTI), Malaysia and Universiti Teknologi Malaysia (UTM).

**Authors**

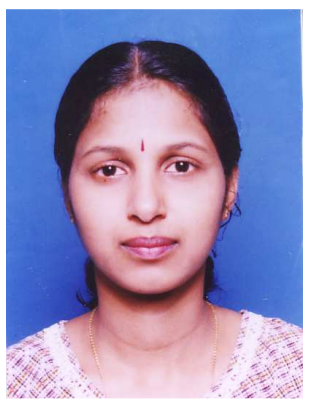

**Kanmani Munusamy** received her Degree in Computer Science from Universiti Teknologi Malaysia (UTM) in 2001 and her Master in Computer Science from Universiti Malaya (UM) in 2004. Currently she is pursuing her PhD in the same field at UTM. Her research interests lie in the areas on Semantic Web Services, process incompatibility analysis and process mediation to support discovery, selection and composition.

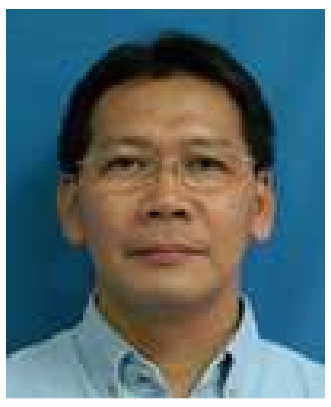

**Harihodin Selamat** received his MSc from University of Cranfield, and his PhD from University of Bradford, United Kingdom. He is currently an Associate Professor in the Faculty of Computer Science and Information System at UTM. He has over 25 years of experience in the field of information systems, system analysis and design. His research interest includes database design and modeling, and enterprise resource planning.

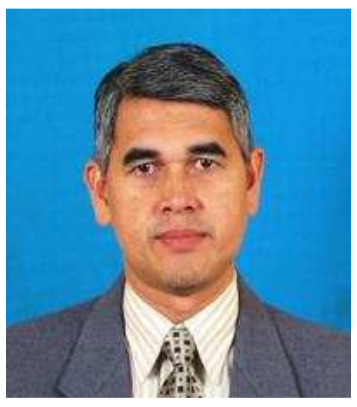

**Suhaimi Ibrahim** received his MSc (1990) and PhD in Computer Science (2006) from UTM, Malaysia. He is an Associate Professor in Advanced Informatics School, UTM and holds the post of Deputy Dean. His research interests include software testing, requirements engineering, Web services, software process improvement, mobile and trusted computing. He also an ISTQB certified tester and being appointed a board member of the Malaysian Software Testing Board (MSTB).

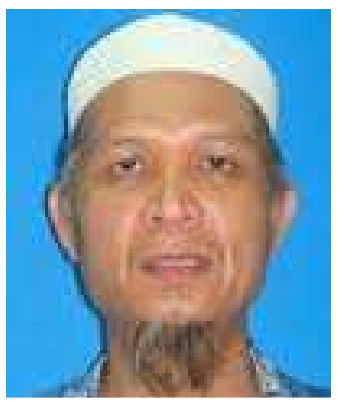

**Sapiyan Baba** received his MSc from Dundee University, Scotland, and his PhD from Keele University, United Kingdom. Currently he is working as Professor in the Department of Artificial Intelligence, Faculty of Computer Science and Information System at UM. His research expertise focuses in Artificial Intelligence applications in various fields and currently he focuses in the field of education.